\documentclass[floats,floatfix,showpacs,amssymb,prd,twocolumn,superscriptaddress,nofootinbib,nolongbibliography,reprint]{revtex4-1}

\usepackage{amssymb,amsmath,verbatim,mathtools,needspace,enumitem,etoolbox,graphicx,physics,microtype,afterpage}

\usepackage[dvipsnames, usenames]{xcolor}

\definecolor{linkcolor}{rgb}{0.0,0.3,0.5}
\usepackage[unicode, colorlinks=true, linkcolor=linkcolor, citecolor=linkcolor, filecolor=linkcolor,urlcolor=linkcolor, pdfusetitle]{hyperref}
\usepackage[all]{hypcap}
\usepackage[T1]{fontenc}
\usepackage[utf8]{inputenc}
\usepackage{tabularx}
\allowdisplaybreaks
\newcommand{\ssim}{\mathchar"5218\relax\,}

\newcommand{\new}[1]{{\textcolor{black}{ #1} }}

\begin{document}

\title{Amplification of superkicks in black-hole binaries through orbital eccentricity}

\author{Ulrich Sperhake}
\email{u.sperhake@damtp.cam.ac.uk}
\affiliation{Department of Applied Mathematics and Theoretical Physics, Centre for Mathematical Sciences, University of Cambridge, Wilberforce Road, Cambridge CB3 0WA, United Kingdom}
\affiliation{Department of Physics and Astronomy, The University of Mississippi, University, Mississippi 38677, USA}
\affiliation{California Institute of Technology, Pasadena, California 91125, USA}

\author{Roxana Rosca-Mead}
\email{rr417@cam.ac.uk}
\affiliation{Department of Applied Mathematics and Theoretical Physics, Centre for Mathematical Sciences, University of Cambridge, Wilberforce Road, Cambridge CB3 0WA, United Kingdom}

\author{Davide Gerosa}
\email{d.gerosa@bham.ac.uk}
\affiliation{School of Physics and Astronomy and Institute for Gravitational Wave Astronomy, University of Birmingham, Birmingham, B15 2TT, United Kingdom}

\author{Emanuele Berti}
\email{berti@jhu.edu}
\affiliation{Department of Physics and Astronomy, Johns Hopkins University, 3400 N. Charles Street, Baltimore, Maryland, 21218, USA}
\affiliation{Department of Physics and Astronomy, The University of Mississippi, University, Mississippi 38677, USA}

%

\date{\today}

\begin{abstract}
  We present new numerical-relativity simulations of eccentric merging
  black holes with initially antiparallel spins lying in the orbital
  plane (the so-called \emph{superkick} configuration). Binary
  eccentricity boosts the recoil of the merger remnant by up to
  $25\%$.
  The increase in the energy flux is much more modest, and therefore
  this kick enhancement is mainly due to asymmetry in the binary
  dynamics.  Our findings might have important consequences for the
  retention of stellar-mass black holes in star clusters and
  supermassive black holes in galactic hosts.
\end{abstract}

\maketitle

\section{Introduction}

According to Einstein's theory of general relativity, gravitational waves carry energy, angular momentum, and linear momentum. In a binary black-hole (BH) system the emission of energy and angular momentum causes the orbit to shrink, eventually leading to the merger of the two BHs. The emission of linear momentum imparts a recoil (or \emph{kick}) to the merger remnant~\cite{1961RSPSA.265..109B,1962PhRv..128.2471P,1973ApJ...183..657B}. 

Calculations based on post-Newtonian (PN) theory found BH recoil
speeds\footnote{Speeds are dimensionless in natural units
($c=G=1$). Therefore, the recoil imparted to a BH does not depend
on the total mass of the system.} of $\mathcal{O}(100)$~km/s
\cite{1983MNRAS.203.1049F,2004ApJ...607L...5F,2005ApJ...635..508B}.
Numerical-relativity (NR) simulations, however, show that BH recoils
can be more than an order of magnitude larger. This is because the
vast majority of the linear momentum is emitted during the last few
orbits and merger, where spin interactions are particularly prominent and
analytic descriptions within the PN framework become inaccurate. In
particular, in 2007 several groups
realized
that binary BHs with spins lying in the orbital plane and
antiparallel to each other might receive \emph{superkicks} as large
as $\ssim 3500$ km/s~\cite{2007PhRvL..98w1101G,2007PhRvL..98w1102C,2007PhRvD..76f1502T}. Subsequent studies found that even larger kicks,
up to $\sim 5000$ km/s, can be reached by further fine-tuning the spin
directions \cite{2011PhRvL.107w1102L,2013PhRvD..87h4027L,Zlochower:2015wga,Lousto:2019lyf}.
 Large kicks strongly affect the dominant mode of
gravitational waveforms
\cite{2008PhRvD..77l4047B,Pretorius:2007nq,2008PhRvD..77d4031S}, and
therefore it should be possible to directly measure their effect with
future gravitational-wave (GW) observations \cite{2016PhRvL.117a1101G,2018PhRvL.121s1102C}.
Further studies targeted hyperbolic encounters \cite{Healy:2008js} and ultrarelativistic collisions (which are not expected to occur in astrophysical settings) \cite{2011PhRvD..83b4037S}, where kicks can reach $10^4$~km/s.
We refer to
Refs.~\cite{2010RvMP...82.3069C,2015ASSP...40..185S,2018PhRvD..97j4049G} for
more extensive reviews on the phenomenology of BH recoils.

The occurrence of superkicks has striking astrophysical consequences
for both stellar-mass and supermassive BHs. In particular, BH recoils
predicted by NR simulations should be compared to the escape speeds of
typical astrophysical environments \cite{2004ApJ...607L...9M}.  

The
stellar-mass BH binaries observed by LIGO and Virgo may form
dynamically in globular clusters \cite{2013LRR....16....4B}, which present
escape velocities in the range $10-50$~km/s.  These values are smaller
even than typical recoil velocities of nonspinning BH binaries
\cite{2007PhRvL..98i1101G}, which implies that a large fraction of
stellar-mass BHs merging in those environments is likely to be ejected
\cite{2018MNRAS.481.2168M} (see Ref.~\cite{2008ApJ...686..829H} for a
complementary study on intermediate-mass BHs in globular clusters). This may
not be the case for environments with larger escape speeds such as
nuclear star clusters \cite{2016ApJ...831..187A} or accretion disks  in active galactic
nuclei \cite{2017MNRAS.464..946S,2017ApJ...835..165B},
which might therefore retain a majority of their merger remnants. If
able to pair again, the BHs in such an environment can form ``second
generation'' GW events detectable by LIGO and Virgo
\cite{2019PhRvD.100d1301G}.

The supermassive BH mergers targeted by LISA and pulsar-timing arrays
(PTAs) may also be significantly affected by large recoils. Superkicks
of $\mathcal{O}(1000)$~km/s exceed the escape speed of even the most
massive elliptical galaxies in our Universe. If supermassive BHs are
efficiently ejected from their galactic hosts, this decreases their
occupation fraction \cite{2015MNRAS.446...38G} and, consequently, LISA
event rates
\cite{2008MNRAS.390.1311B,2007MNRAS.382L...6S}. Spin-alignment
processes of both astrophysical
\cite{2007ApJ...661L.147B,2013MNRAS.429L..30L,2013ApJ...774...43M,2015MNRAS.451.3941G}
and relativistic \cite{2010ApJ...715.1006K,2012PhRvD..85l4049B} nature
are commonly invoked to mitigate this effect.

Recoils are driven by asymmetries in the merging binary
\cite{2008PhRvL.100o1101B,2008PhRvD..78b4017B}; no kick can be
imparted if the emission of gravitational-wave energy is isotropic.
\begin{figure*}[t]
  \includegraphics[width=0.48\textwidth]{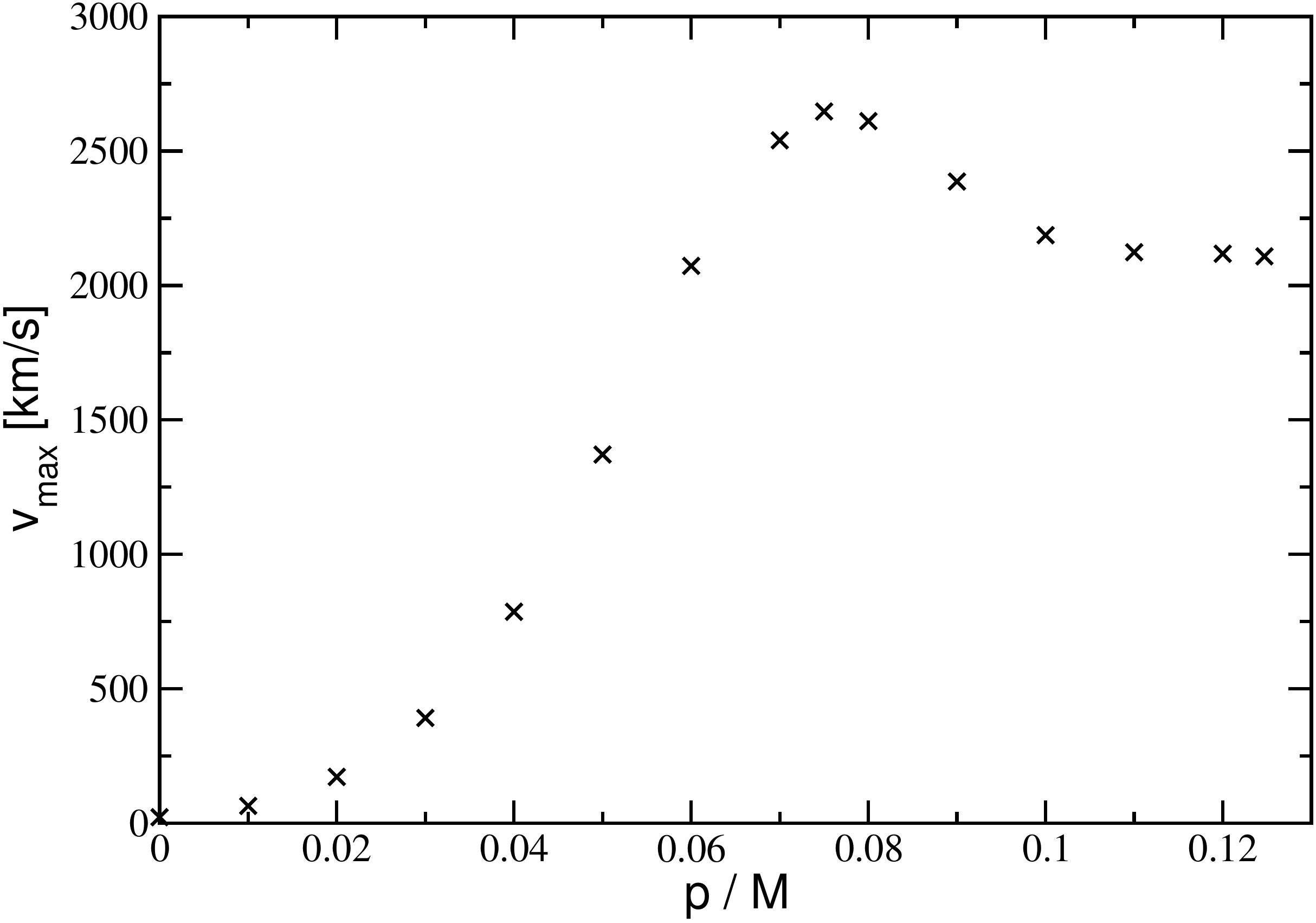}
  \hspace{0.5cm}
  \includegraphics[width=0.48\textwidth]{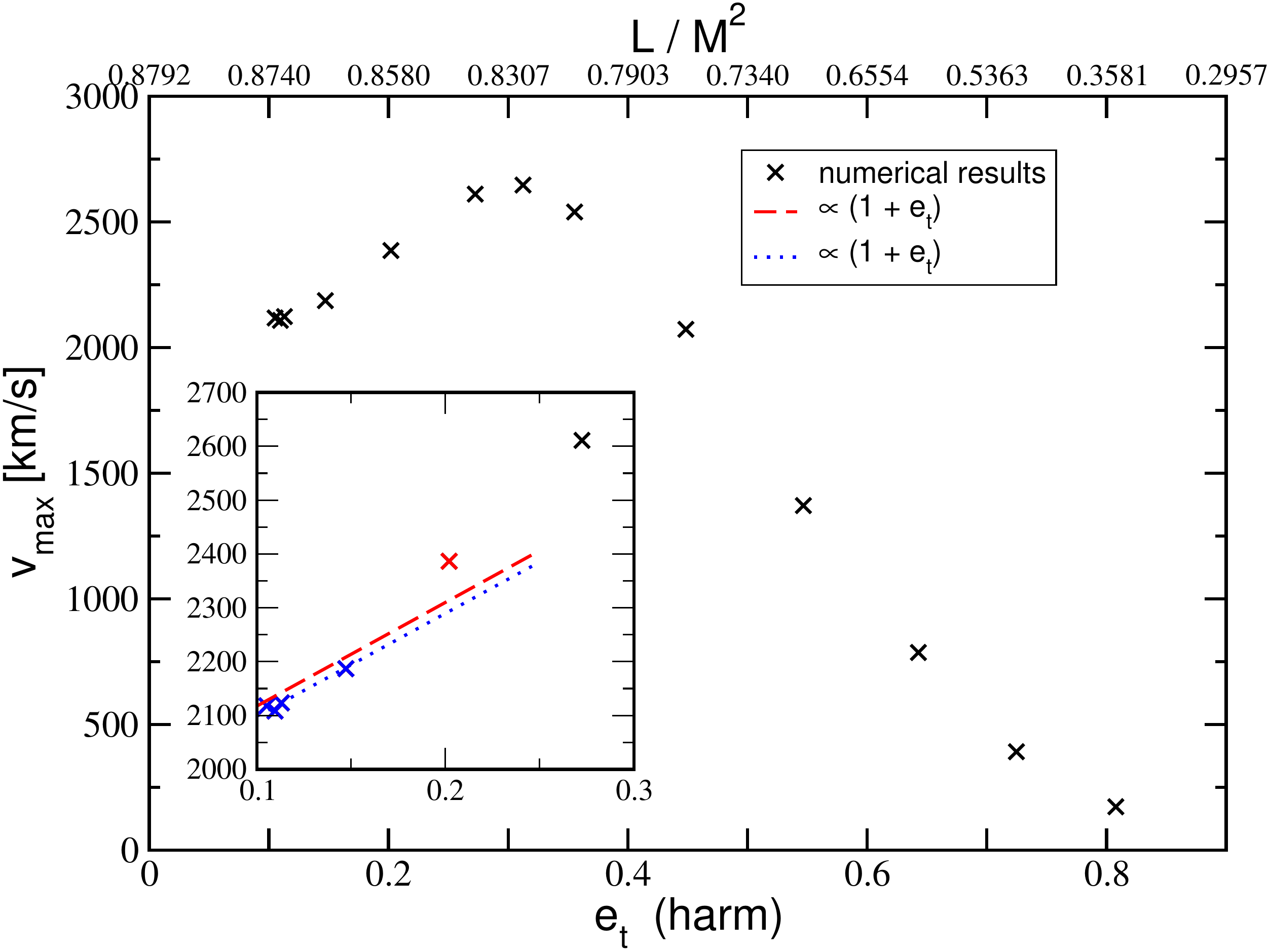}
  \caption{Superkicks for eccentric binary BHs with equal masses and spins of magnitude $\chi_1=\chi_2=0.596\simeq 0.6$. {\em Left}:
           The maximum kick velocity $v_{\rm max}$ as a function of the
           linear momentum parameter.
           The largest kicks correspond to moderate eccentricity and exceed the quasicircular value
           by about $25\,\%$.
           {\em Right}:
           The maximum kick velocity $v_{\rm max}$ as a function of the
           eccentricity parameter $e_t$ estimated in harmonic gauge.
           Labels on the upper horizontal axis display the corresponding
           initial orbital angular momentum $L/M^2$ of the binaries.
           The inset zooms in on the low-eccentricity regime and
           shows linear fits $v_{\rm kick}\propto (1+e_t)$
           obtained from the first four, blue data points
           (dotted curve), and also including the fifth, red data point
           (dashed curve). The increase of the recoil for small eccentricity
           is compatible with the $(1+e)$ scaling from close-limit
           calculations \cite{2007ApJ...656L...9S}.
          }
  \label{fig:ampkickofp}
\end{figure*}
For instance, an equal-mass nonspinning binary does not
recoil by symmetry. Unequal masses or
misaligned spins, however, introduce
asymmetries in the GW emission. Orbital eccentricity is a
further natural ingredient to enhance the asymmetry of the binary and,
consequently, the kick. Early PN estimates show that, for low eccentricities
$e\lesssim 0.1$, the kick imparted to nonspinning BHs increases by
about $10\%$,
with a scaling proportional to $1+e$
\cite{2007ApJ...656L...9S}.

In this paper, we investigate for the first time how superkicks are
affected by binary eccentricity using NR simulations of the merger.
For this purpose, we consider equal-mass binaries with
$M_1=M_2\equiv M/2$ with BH spins of equal magnitude pointing in
opposite directions inside the orbital plane,
$\boldsymbol{S}_1=-\boldsymbol{S}_2$.  Fixing the dimensionless spin
$\chi_{i}\equiv |\boldsymbol{S}_{i}|/M_{i}^2$ to $\chi_1=\chi_2=0.596$,
we generate a sequence of increasing
eccentricity by gradually reducing the initial orbital angular
momentum $L$ at fixed binding energy from the quasicircular value to
the head-on limit $L=0$; in practice, we vary for this purpose the
initial tangential momentum parameter $p$ of each BH.  For a given
eccentricity (i.e.~fixed $L$), the kick is known to depend
sinusoidally on the initial angle of the two spins relative to the
line connecting the BHs
\cite{Kidder:1995zr,Campanelli:2007cga,2008PhRvD..77l4047B}.
The maximum value of this sine function is
the kick reported in Fig.~\ref{fig:ampkickofp} as a function of the
linear momentum and of the eccentricity. The significant increase of
the maximum kick from about $2100$~km/s for approximately quasicircular
binaries to $2600$~km/s for moderate eccentricities
$e_t\ssim 0.3$ is the main finding of our study (where $e_t$ is the eccentricity parameter of  Ref.~\cite{2008PhRvD..78f4069S,Memmesheimer:2004cv}). We furthermore show
that such an increase holds over a wider range of spin magnitudes and
correspondingly raises the maximum superkick in BH binaries to about
$4200$~km/s, larger than the maximum of $\ssim 3700$~km/s for
negligible eccentricity.

The rest of this paper presents our methodology and results in more detail and
is organized as follows. In Sec.~\ref{secNR},
we describe our NR runs; in Sec.~\ref{secresults}, we present our
recoil analysis; and in Sec.~\ref{secconcl}, we discuss the astrophysical
relevance of our findings and possible directions for future work.

\section{Computational framework and set of simulations}
\label{secNR}
\subsection{Numerical-relativity setup}

The BH binary simulations reported in this work have been
performed with the {\sc Lean} code \cite{2007PhRvD..76j4015S},
which is based on the {\sc Cactus} computational toolkit
\cite{Allen:1999,Cactusweb}. The Einstein equations
are implemented in the form of the
Baumgarte-Shapiro-Shibata-Nakamura-Oohara-Kojima (BSSNOK) formulation
\cite{Nakamura:1987zz,Shibata:1995we,1999PhRvD..59b4007B} using
the method of lines with fourth-order Runge-Kutta differencing
in time and sixth-order stencils in space for improved phase
accuracy \cite{2008CQGra..25j5006H}. The wide range of length scales
is accommodated through adaptive mesh refinement provided by
{\sc carpet} \cite{2004CQGra..21.1465S,Carpetweb}
and we compute apparent horizons with
{\sc ahfinderdirect} \cite{1996PhRvD..54.4899T,2004CQGra..21..743T}.
We start our simulations with puncture \cite{1997PhRvL..78.3606B}
data of Bowen-York \cite{Bowen:1980yu} type
computed with Ansorg's spectral solver \cite{2004PhRvD..70f4011A}
inside the {\sc cactus} {\sc twopuncture} thorn
and evolve these using the moving puncture approach
\cite{Baker:2005vv,Campanelli:2005dd}.
The gravitational wave signal is extracted in the form
of the Newman-Penrose scalar $\Psi_4$ computed from the
grid variables
~\cite{2007PhRvD..76j4015S}.

\subsection{Black-hole binary configurations}
In this study, we consider equal-mass BH binaries in the
superkick configuration; i.e.~the BHs have spins of equal magnitude
pointing in opposite directions in the orbital plane.\footnote{We
define here the orbital plane as the plane spanned by the initial
position vector connecting the BHs and their initial linear momentum
--- in our case this is the $xy$ plane, and the $z$ axis points in
the direction perpendicular to this plane.}
In practice, we do not compute the dimensionless spins $\chi_i$
directly from the Bowen-York spin, because some angular momentum and
energy are contained in the spurious radiation of the conformally flat
initial data.  This energy and momentum are partly accreted onto the
BHs and partly radiated to infinity, leading to a brief period of spin
adjustment.  While negligible for slowly rotating BHs, this effect
increases for larger spin parameters and ultimately leads to a
saturation at $\chi \ssim 0.928$ \cite{Cook:1989fb,2002PhRvD..65j4038D}.  In
order to obtain a more accurate estimate of $\chi_i$, we monitor the
BH spins $\boldsymbol{S}_i$ using the method described in
Ref.~\cite{2007PhRvD..75f4030C} and compute the irreducible mass
$m_{\rm ir}$ from the apparent horizon during the evolution.  The
dimensionless spin $\chi_i$ can then be computed according to
\cite{Christodoulou:1970wf}
\begin{equation}
  M_i^2 = m_{{\rm ir},i}^2 + \frac{|\boldsymbol{S}_i|^2}{4m_{{\rm ir},i}^2}\,.
        ~~~~~
  \chi_i = \frac{|\boldsymbol{S}_i|}{M_i^2}\,.
\end{equation}
As expected from the above description, we observe a brief transient period
in all simulations
during which $\chi_i$ mildly decreases. Throughout this work we report the
initial spin as the value at time $t_{\chi}=20M$
measured from the beginning of the simulation. By this
time $\chi_i$ has reached a nearly stationary value, so that the precise
value of $t_{\chi}$ does not affect the results.
We distinguish this estimate for the initial spin from the
value directly obtained from the Bowen-York parameters, which we denote by
$\chi_{{\rm BY},i}$. The relation between $\chi_i$ and $\chi_{{\rm BY},i}$
is shown in the fourth and fifth columns of
Table \ref{tab:short}.
All simulations presented in this paper have $\chi_1=\chi_2$.

The net spin is zero in the superkick configurations, resulting in
dynamics rather similar to those of nonspinning BH binaries; the
main difference is a periodic motion of the orbital plane in the
orthogonal (in our case $z$) direction.  This motion of the binary
orthogonal to the orbital plane results in a periodic blue- and redshift
of the gravitational radiation and the net effect of this
beaming leads to asymmetric GW emission, especially in the
$(\ell,m)=(2,2)$ and $(2,-2)$ multipoles and, hence, net emission
of linear momentum and the ensuing
recoil of the postmerger remnant \cite{2008PhRvD..77l4047B,Pretorius:2007nq}.
For fixed initial position $(\pm x_0,\,0,\,0)$ of the BH binary,
the periodic nature of the blue- and redshifting of the gravitational
radiation furthermore manifests itself in a sinusoidal dependence of
the actual kick magnitude on the initial orientation of the spins
in the orbital plane \cite{2007ApJ...659L...5C,2008PhRvD..77l4047B}.
We quantify this orientation in terms of the angle $\alpha$
between the initial spin of the BH starting at $x>0$ and the
$x$ axis; i.e.~this BH has initial spin $\boldsymbol{S}_1
=S\,(\cos \alpha,	\,\sin\alpha,\,0)$, while the BH at $x<0$
is initialized with $\boldsymbol{S}_2=-\boldsymbol{S}_1$  \cite{2008PhRvD..77l4047B,2018PhRvD..97j4049G}.

In order to assess the impact of the orbital eccentricity on the
magnitude of the gravitational recoil, we have constructed a set
of binary configurations guided by the second sequence
of equal-mass, nonspinning
BH binaries in Table I of Ref.~\cite{2008PhRvD..78f4069S}. This sequence
starts with a quasicircular binary with initial separation $D/M=7$
and a tangential linear momentum $p/M=0.1247$ for each BH, resulting
in an orbital angular momentum $L/M^2=0.8729$.
These parameters determine
the binding energy of the binary through $E_{\rm b}\equiv M_{\rm ADM}-M$,
where $M_{\rm ADM}$ is the Arnowitt-Deser-Misner (ADM) mass
\cite{Arnowitt:1962hi} of the binary spacetime. We construct a sequence
of configurations with increasing eccentricity by gradually
reducing the initial linear momentum parameter while keeping the
binding energy fixed at $E_b/M=-0.012$.
For this choice, the gradual reduction of initial kinetic
energy for larger eccentricity
implies a larger initial separation, i.e.~correspondingly
less negative potential energy, and, thus, ensures an inspiral
phase of comparable duration irrespective of the eccentricity.

The variation in the initial separation of the BHs requires a minor change
in the setup of the computational grid for low- and high-eccentricity
binaries. In the notation
of Ref.~\cite{2007PhRvD..76j4015S} we
employ a grid setup given in units of $M$ by
\begin{eqnarray}
  && \{(256,\,128,\,64,\,32,\,16,\,8) \times (2,\,1),~h\}\,,
        \nonumber \\[5pt]
  && \{(256,\,128,\,64,\,32,\,16)\times(4,\,2,\,1),~h\}\,,
  \label{eq:grid}
\end{eqnarray}
respectively, for binaries with $p/M\ge 0.8$ and those with $p/M<0.8$.
\new{
Here, the first line specifies a computational domain with six fixed outer
grid components of cubic shape centered on the origin
with radius 256, 128, 64, 32, 16, and 8,
respectively, and two refinement levels with two cubic components each with
radius 2 and 1 centered around either hole. The grid spacing is $h$ on the
innermost level and successively increases by a factor of 2 on each
next outer level. The second line in (\ref{eq:grid})
likewise specifies a grid with five
fixed and three dynamic refinement levels.
}
Unless stated otherwise, we use a resolution $h=M/64$.

In order to accommodate the above-mentioned sinusoidal variation
of the kick velocity with the initial spin orientation $\alpha$,
we have performed for each value of the linear momentum parameter $p$
a subset of 6 runs with $\alpha \in [0,\,180^{\circ})$. Due to the symmetry
of the superkick configuration under a shift of the azimuthal angle
$\phi\rightarrow \phi + 180^{\circ}$, the recoil will always point in the
$z$ direction with $v_x=v_y=0$ \cite{2008PhRvL.100o1101B,2008PhRvD..78b4017B}. Furthermore, two binaries with
initial spin orientations $\alpha$ and $\alpha+180^{\circ}$ will generate kicks
of equal magnitude but opposite direction, i.e.~$v_z(\alpha)=
-v_z(\alpha+180^{\circ})$
\cite{2008PhRvD..77l4047B}.
Kick velocities for $\alpha \ge 180^{\circ}$ can therefore be directly inferred
through this symmetry from the simulations performed.
For a few selected cases, we
have performed additional simulations with $\alpha\ge 180^{\circ}$;
the symmetry is confirmed with accuracy of $\mathcal{O}(0.1)~\%$
or better.

\subsection{Measuring the eccentricity}

Our sequence of simulations is characterized by the variation of the
orbital angular momentum at fixed binding energy. As discussed in
detail in Ref.~\cite{2008PhRvD..78f4069S}, there is no unambiguous way
to assign an eccentricity parameter to BH binaries in the late stages
of the inspiral. Motivated by the close similarity of the orbital
dynamics of (equal-mass) superkick binaries and nonspinning binaries,
we follow here the procedure used in Ref.~\cite{2008PhRvD..78f4069S} to
obtain a PN estimate for nonspinning binaries. Specifically, we use
Eqs.~(20) and (25) of Ref.~\cite{Memmesheimer:2004cv}, which provide the
PN eccentricity parameter $e_t$ for nonspinning binaries. This
estimate needs to be taken with a grain of salt as it is only an
approximation at the small binary separation during the last orbits
before merger, and it ignores the effect of BH spins. Furthermore
$e_t$ exhibits an infinite gradient near the quasicircular limit when
plotted as a function of the orbital angular momentum, leading to
limited precision for values $e_t\lesssim 0.1$. Similarly, in the
head-on limit the vanishing of $L$ leads to a formal divergence of the
eccentricity parameter and a Newtonian interpretation ceases to be
valid (values $e_t>1$ are possible in this regime).  Nevertheless,
$e_t$ provides us with a rough estimate to quantify deviations from
the quasicircular case and distinguish low-, moderate- and
high-eccentricity configurations.

For all simulations, we have computed the following diagnostic
variables. The energy, linear and angular momentum radiated
in GWs are computed on extraction spheres
of coordinate radius $r_{\rm ex}/M=30,\,40,\,\ldots,\,90$
from the Newman-Penrose scalar
according to the standard methods described, for example, in
Ref.~\cite{2008GReGr..40.1705R}.
For the physical radiation reported in Table
\ref{tab:short},
we exclude the spurious radiation inherent in
the initial data by considering only the wave signal starting at
retarded time $u\equiv t-r_{\rm ex}=50~M$.
We also compute the dimensionless spin
of the postmerger BH from the apparent horizon \cite{2018CQGra..35w5008C}.
We have confirmed these values using also the conservation of energy
and angular momentum, which yields agreement to within $0.5~\%$ or better.

\subsection{Numerical accuracy}
Our numerical results for the GW emission and the recoil velocities
are affected by two main sources of uncertainty: the discretization
error and the finite extraction radii for the Newman-Penrose scalar.

We address the latter by extrapolating the GW signal to infinity
using a Taylor series in $1/r$ 
as in Ref.~\cite{2011CQGra..28m4004S}. The results reported are those extrapolated
at linear order in $1/r$, and we estimate the error through the difference
with respect to a second-order extrapolation. The magnitude of this error is
$\ssim 2\,\%$ or less.

In order to assess the error due to finite differencing, we
have performed additional simulations
of the configuration $p/M=0.1247$, $\chi_i=0.596$, $\alpha=150^{\circ}$
using grid resolutions
$h=M/48$ and $h=M/80$.
\begin{figure}[t]
  \includegraphics[width=0.48\textwidth]{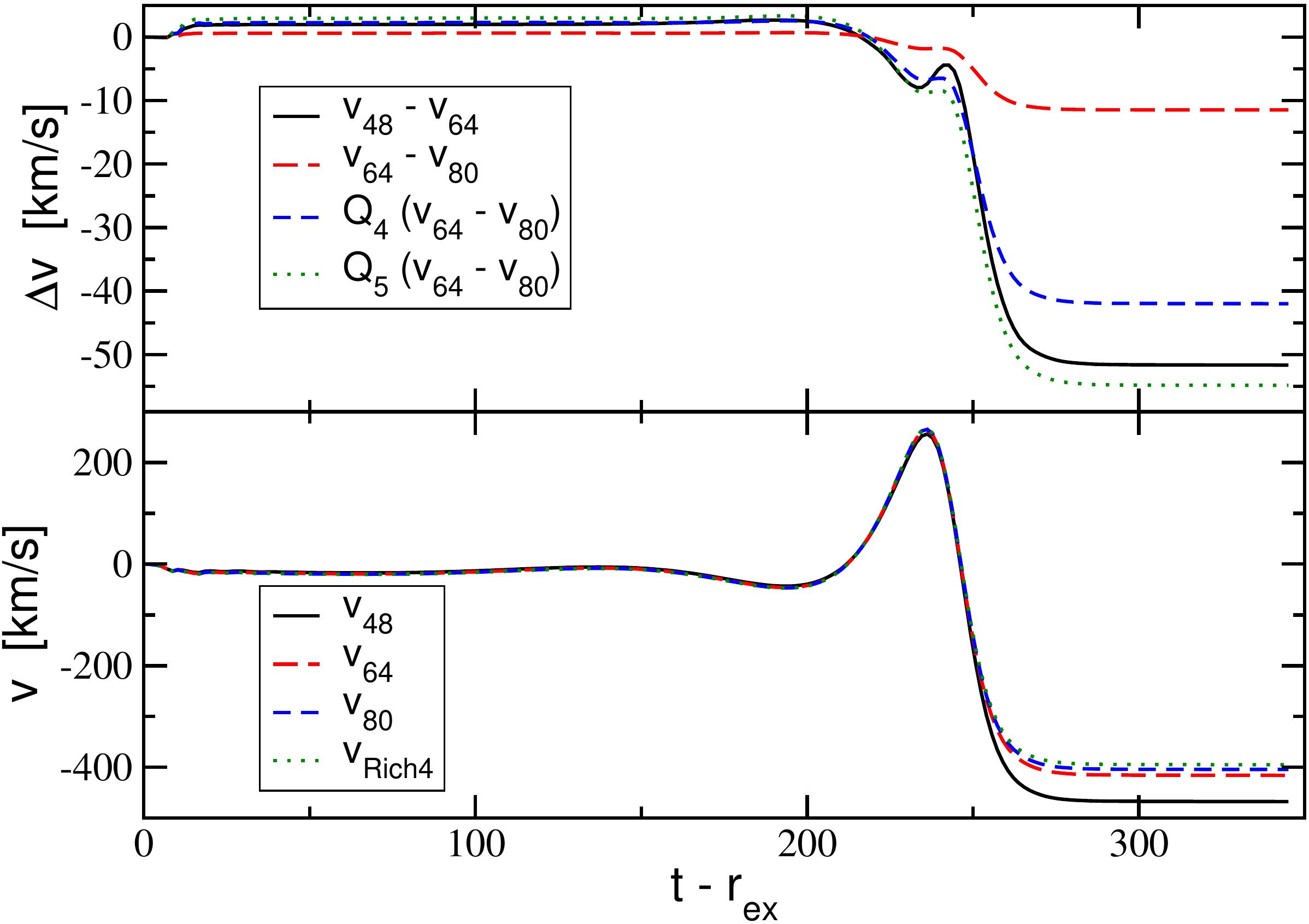}
  \caption{Convergence analysis for the linear momentum radiated from
    a binary with $p/M=0.1247$, $\chi_i=0.596$, $\alpha=150^{\circ}$.  The
    linear momentum obtained for resolutions $h=M/48$, $h=M/64$, and
    $h=M/80$ is shown in the form of the kick velocity accumulated up
    to retarded time $u=t-r_{\rm ex}$ in the bottom panel. The upper
    panel shows differences between various resolutions, together with
    rescaling according to fourth- and fifth-order convergence.  We
    estimate the uncertainty from the more conservative fourth-order
    Richardson extrapolation, and we obtain a numerical error estimate
    of about $2\,\%$ for our standard resolution $h=M/64$.  }
  \label{fig:convP}
\end{figure}
Figure~\ref{fig:convP} shows convergence between
fourth and fifth order resulting in a discretization error of about $2\,\%$
for the radiated linear momentum. A similar behavior is observed for
the radiated energy $E_{\rm rad}$.
We use this value as an error estimate, but note
that this is a conservative estimate for
the {\em maximum} kick velocity at fixed eccentricity.
The reason is that a considerable part of the numerical error consists in
the inaccuracy of the inspiral phase of the binary.
This phase error significantly affects the angle
$\alpha_0$ in Eq.~(\ref{eq:vfit}) below, but has weaker repercussions on
the maximum kick $v_{\rm max}$.
In other words, at lower resolution, we will
obtain the maximum kick at a ``wrong'' phase angle $\alpha_0$, but still
measure this maximum with decent precision. We have verified this
expectation by generating a complete sequence for $p/M=0.1247$, $\chi_i=0.596$
at low, medium and high resolution. Applying the fit (\ref{eq:vfit}) to
each of these gives us
$v_{\rm max}=2098.1, 2108.3,$ and $2109.7~{\rm km/s}$, respectively, for $h/M=1/48$, $1/64$, and
$1/80$. Since we cannot entirely rule out fortuitous cancellation of
errors
in this excellent agreement, we keep in the remainder of this work the
more conservative $2\,\%$ estimate from Fig.~\ref{fig:convP}.
Combined with the extrapolation procedure to $r_{\rm ex}\rightarrow
\infty$, we estimate our total error budget as $\sim 4\,\%$.

\section{Numerical results}
\label{secresults}

%

The main results of our study are summarized in Table
\ref{tab:short}. For each sequence with prescribed linear momentum
$p$, we list there the initial separation $D$,
orbital angular momentum $L$, the initial BH spins $\chi_{{\rm BY},i}$ and
$\chi_i$, eccentricity estimates $e_t$ obtained in ADMTT and harmonic gauge
according to Eqs.~(20) and (25) of Ref.~\cite{Memmesheimer:2004cv},
the mean radiated energy $E_0$,
the maximum kick velocity $v_{\rm max}$,
and the dimensionless spin $\chi_0$ of the merger remnant.
\begin{table*}
\begin{tabular}{lcccc|ccccc}
\hline \hline
$p/M$~~~&~~~$D/M$~~~&~~~$L/M^2$~~~&$\chi_{{\rm BY},1}=\chi_{{\rm BY},2}$&~~~$\chi_1=\chi_2$~~~&~~~$e_t$(ADMTT)~~~&~~~$e_t$(harm)~~~&~~~$10^2\,E_0/M$~~~&~~~$v_{\rm max}~[{\rm km/s}]$~~~&~~~$\chi_0$~~~
\\
\hline
\hline
0.1247& 7.000& 0.8729 &0.6 &0.596& 0.1095 & 0.1096 & 3.687 & 2108 & 0.6815 \\
0.12  & 7.278& 0.8734 &0.6 &0.596& 0.1049 & 0.1052 & 3.678 & 2118 & 0.6810 \\
0.11  & 7.932& 0.8725 &0.6 &0.596& 0.1130 & 0.1130 & 3.664 & 2123 & 0.6798 \\
0.10  & 8.678& 0.8678 &0.6 &0.596& 0.1480 & 0.1472 & 3.757 & 2187 & 0.6808 \\
0.09  & 9.529& 0.8576 &0.6 &0.596& 0.2040 & 0.2020 & 3.862 & 2387 & 0.6884 \\
0.08  &10.493& 0.8394 &0.6 &0.596& 0.2758 & 0.2725 & 3.656 & 2611 & 0.6999 \\
0.075 &11.018& 0.8264 &0.6 &0.596& 0.3166 & 0.3124 & 3.368 & 2647 & 0.7010 \\
0.07  &11.571& 0.8100 &0.6 &0.596& 0.3608 & 0.3555 & 3.069 & 2540 & 0.7021 \\
0.06  &12.754& 0.7652 &0.6 &0.596& 0.4567 & 0.4485 & 2.258 & 2073 & 0.6905 \\
0.05  &14.013& 0.7007 &0.6 &0.596& 0.5603 & 0.5467 & 1.452 & 1371 & 0.6539 \\
0.04  &15.288& 0.6115 &0.6 &0.596& 0.6681 & 0.6428 & 0.833 &  786 & 0.5862 \\
0.03  &16.487& 0.4946 &0.6 &0.596& 0.7835 & 0.7247 & 0.429 &  391 & 0.4839 \\
0.02  &17.488& 0.3498 &0.6 &0.596& 1.0122 & 0.8078 & 0.203 &  172 & 0.3467 \\
0.01  &18.162& 0.1816 &0.6 &0.596& 3.0771 & 2.0975 & 0.100 &   64 & 0.1813 \\
0     &18.398& 0      &0.6 &0.596&$\infty$&$\infty$& 0.071 &   22 & 0      \\
\hline
0.075 &11.018& 0.8264 &0.6 &0.596& 0.3166 & 0.3124 & 3.368 & 2647 & 0.7010 \\
0.075 &11.018& 0.8264 &0.65 &0.645& 0.3166 & 0.3124 & 3.383 & 2849 & 0.7002 \\
0.075 &11.018& 0.8264 &0.7 &0.694& 0.3166 & 0.3124 & 3.368 & 3019 & 0.6990 \\
0.075 &11.018& 0.8264 &0.75 &0.742& 0.3166 & 0.3124 & 3.386 & 3166 & 0.6969 \\
0.075 &11.018& 0.8264 &0.8 &0.789& 0.3166 & 0.3124 & 3.330 & 3479 & 0.6976 \\
0.075 &11.018& 0.8264 &0.85 &0.834& 0.3166 & 0.3124 & 3.233 & 3583 & 0.6960 \\
0.075 &11.018& 0.8264 &0.9 &0.876& 0.3166 & 0.3124 & 3.167 & 3776 & 0.6950 \\
\hline
\hline
\end{tabular}
\caption{Each sequence of simulations is characterized by the linear momentum
         parameter $p$ and the initial BH separation $D$ (which determine
         the orbital angular momentum $L$ and the eccentricity of the binary), as well as the initial spins, given here
         in both the form of the pristine Bowen-York parameters $\chi_{{\rm BY},i}$ and of the
         more accurate horizon estimate $\chi_i$.
         The remaining columns list: estimates of the eccentricity $e_t$
         obtained from PN relations in the ADMTT and harmonic gauge,
         respectively; the mean radiated GW energy $E_0$;
         the maximum kick velocity $v_{\rm max} $; and
         the mean spin $\chi_0$ of the remnant BH.
       }
\label{tab:short}
\end{table*}

\subsection{Impact of the orbital eccentricity}

\begin{figure}[t]
  \includegraphics[width=0.48\textwidth]{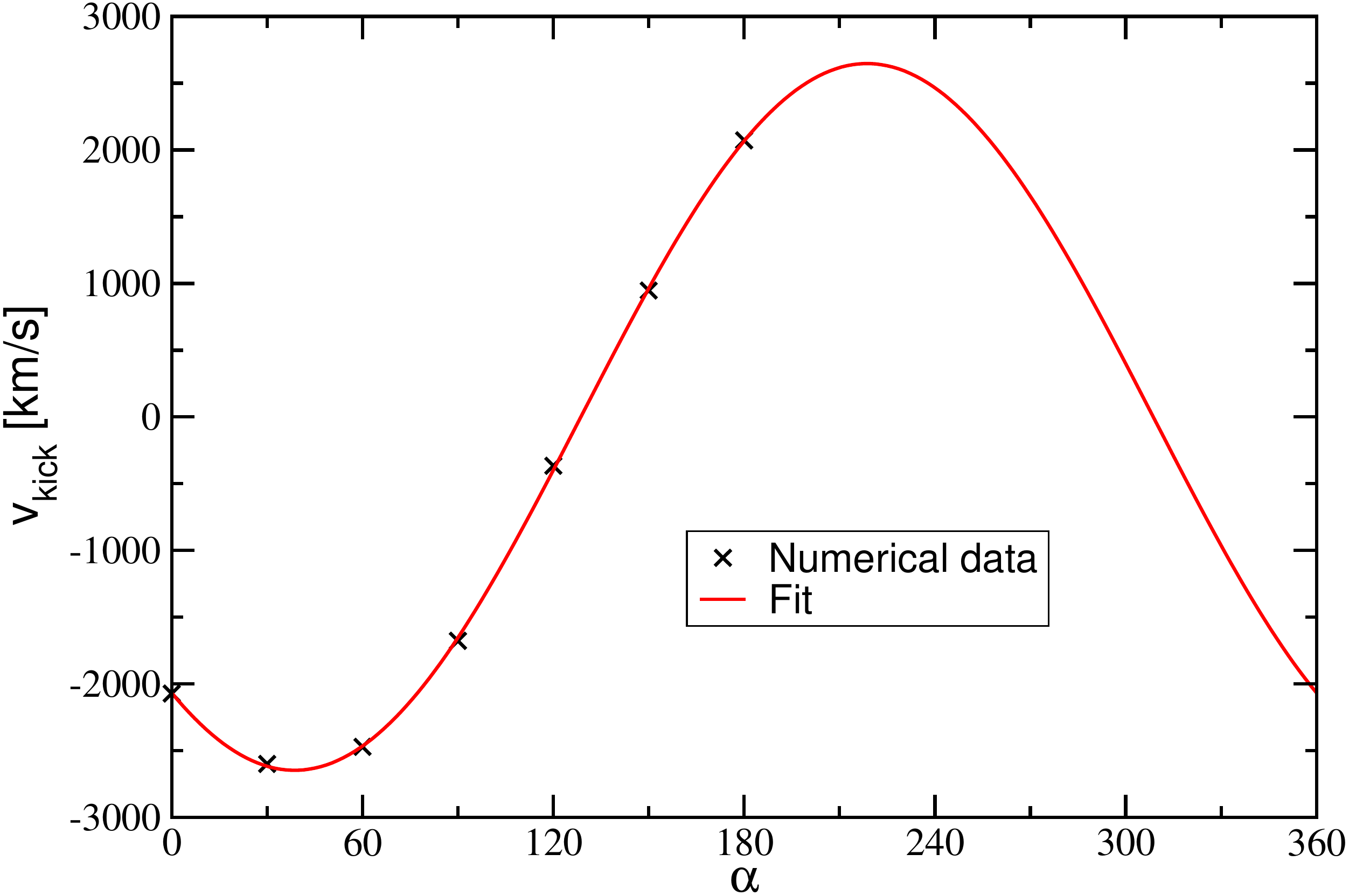}
  \caption{The recoil velocity computed numerically from the
           GW signal for $p=0.075$ is shown as $\times$ symbols
           for different values of the spin orientation $\alpha$.
           The red curve represents the fit obtained from
           these data according to Eq.~(\ref{eq:vfit}).
           Only data for $\alpha<180^{\circ}$ are
           needed to determine the fitting coefficients, due to the
           symmetry of the superkick configurations.
         }
  \label{fig:vofa_p075}
\end{figure}

The sinusoidal dependence of the kick magnitude on the initial
spin orientation $\alpha$ is
illustrated in Fig.~\ref{fig:vofa_p075} for the case
$p/M=0.075$, $\chi_i=0.596$.
The data are reproduced with
high precision by a fit of the form
\begin{equation}
  v_{\rm kick} = v_{\rm max}\times \cos(\alpha-\alpha_0)\,,
        \label{eq:vfit}
\end{equation}
where, for this specific series,
$v_{\rm max}=2647~{\rm km/s}$ and $\alpha_0=218.7^{\circ}$.
The radiated energy $E_{\rm rad}$ and the final spin, in contrast,
vary only mildly (within the numerical uncertainties)
with the angle $\alpha$; we report average values for
these quantities.
More specifically, we fit $E_{\rm rad} =  E_0+ E_1 \sin(2\alpha+\alpha_0)$ and report $E_0$ (and likewise $\chi_0$).

The variation of the kick velocity with eccentricity is visualized
in the left panel of Fig.~\ref{fig:ampkickofp},
which shows $v_{\rm max}$ as a function of the linear momentum $p$.
We clearly see that the largest kicks are not realized for quasicircular
binaries but for moderate eccentricities. A similar effect is apparent
for the radiated energy values of Table \ref{tab:short}, which closely
resembles the observation in Table I of Ref.~\cite{2008PhRvD..78f4069S}
for the nonspinning case. The increase in the recoil velocity, however,
is much stronger: for $p/M=0.075$, the maximum kick exceeds the quasicircular
value by about $25\,\%$, while the largest energy represents a meager
$5\,\%$ increase relative to the quasicircular case. This discrepancy
shows that the enhanced kick is not merely due to increased radiation,
but also to a higher degree of asymmetry in eccentric binaries.

An increase in the recoil at small eccentricities has already been noticed
in the close-limit calculations of Refs.~\cite{2007ApJ...656L...9S,Sopuerta:2006wj},
which find a $(1+e)$ proportionality for eccentricities $e\lesssim 0.1$.
In the right panel of
Fig.~\ref{fig:ampkickofp}, we plot the maximum kick velocity as
a function of the eccentricity parameter $e_t$ in harmonic gauge (the ADMTT
version of $e_t$ would result in virtually the same figure).
Due to the diverging gradient of $e_t$ with respect to the orbital
angular momentum \cite{2008PhRvD..78f4069S}, our data points are limited
to $e_t\gtrsim 0.1$, but as shown in the inset of the figure, the data
are compatible with the linear growth $\propto (1+e_t)$ of the close-limit
approximation. The two fits shown in the inset have been obtained
using either the first four or the first five data points with the expression
$v_{\rm max}=v_0(1+e_t)$. The numerical results suggest that above
$e_t\approx 0.2$, $v_{\rm max}$ increases even more strongly with $e_t$
before reaching the maximum at $e_t\approx 0.3$, and then decreases for
yet higher eccentricity.

\subsection{Impact of the spin magnitudes}

The gravitational recoil in superkick configurations is known to increase
approximately linearly with the spin magnitudes $\chi_i$.
Extrapolating numerical results to maximal spin $\chi_i=1$
results in a maximal superkick of about $3680~{\rm km/s}$
\cite{2007ApJ...659L...5C} for quasicircular binaries. We will now
investigate to what extent nonzero eccentricity can increase this
upper limit.
In order to keep the computational costs manageable, we focus for this
purpose on
the $p/M=0.075$ sequence which maximizes the recoil in our eccentricity
analysis for
$\chi_i=0.596$. We cannot rule out that the ``optimal'' eccentricity
maximizing recoil depends on the spin magnitude, so our analysis should
be regarded as a conservative estimate; the largest possible
superkick in eccentric binaries
may even exceed the value resulting from the analysis below.

We vary the initial spin magnitude $\chi_i$ while keeping
all other parameters, including the eccentricity $e_t$, fixed.
A convergence analysis for $\chi_i=0.9$ yields a similar order as in
Fig.~\ref{fig:convP}, but demonstrates that higher resolution is needed for
these configurations. We use $h=M/80$ for the simulations discussed in this
subsection, which results in a discretization error of about $4\,\%$.
As before, we cover the range of the
initial spin orientation by evolving
six binaries with $\alpha \in [0,180^{\circ})$
for each value of $\chi_i$ and fit the resulting $v_{\rm kick}$
according to the sinusoidal function of Eq.~(\ref{eq:vfit}).
The results for these simulations are listed in the lower block
of Table \ref{tab:short}.
As expected, the maximum recoil velocity $v_{\rm max}$
increases with the spins $\chi_i$.
\begin{figure}
  \includegraphics[width=0.48\textwidth]{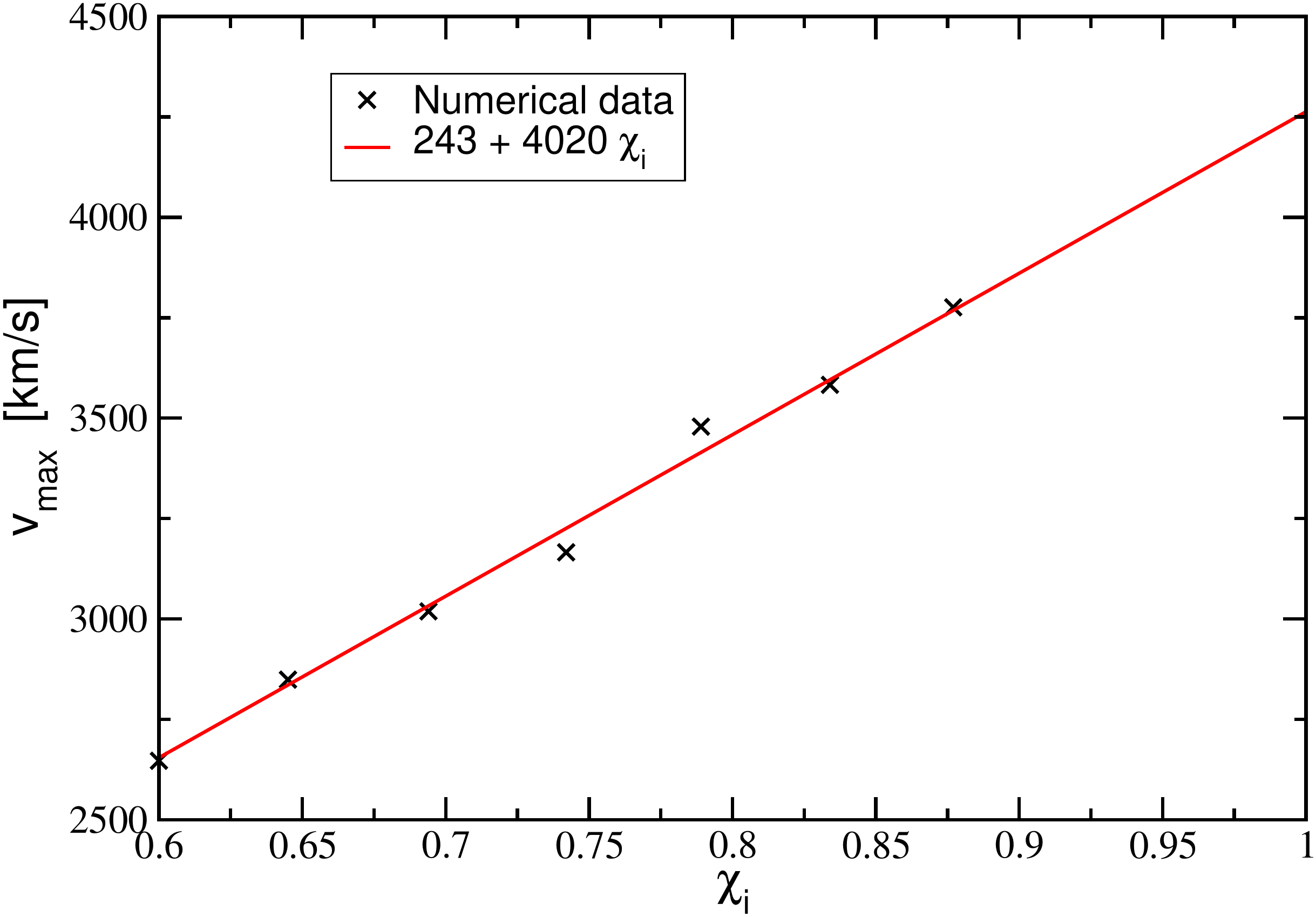}
  \caption{The maximum recoil velocity $v_{\rm max}$ for $p/M=0.075$
           as a function of the initial spin magnitude $\chi_i$.
           The curve represents the linear fit (\ref{eq:maxvofj}).
          }
  \label{fig:maxvofj}
\end{figure}
We display $v_{\rm max}$
as a function of $\chi_i$ in Fig.~\ref{fig:maxvofj},
together with a linear fit to model the leading-order dependence
of the maximum recoil velocity $v_{\rm max}$ on the spin magnitude
$\chi_i$ \cite{2007ApJ...659L...5C,2007PhRvL..98w1102C}.
This fit is given by
\begin{equation}
  v_{\rm max} = \big[ (243\pm 122)~+~(4020\pm 163)\,\chi_i\big]~{\rm km/s}
        \label{eq:maxvofj}
\end{equation}
and predicts a maximum kick of $4263\pm 285~{\rm km/s}$ for extremal spins
$\chi_i=1$. This value exceeds the maximal superkick for quasicircular
binaries of about $3680~{\rm km/s}$
\cite{2007ApJ...659L...5C,2007PhRvL..98w1102C} by about $16\,\%$, but falls
short of the $5000~{\rm km/s}$ maximum for the hang-up kicks reported in
Ref.~\cite{2011PhRvL.107w1102L}. To the best of our knowledge, the effect of eccentricity on these hang-up
kicks has, not yet been explored.
The results reported here and the findings of
Ref.~\cite{2007ApJ...656L...9S} hint that yet larger recoils may be possible
in bound BH binary systems.

\section{Conclusions}
\label{secconcl}

Orbital eccentricity amplifies superkicks. We have presented an extensive
series of numerical simulations of merging BHs with spin vectors of
magnitude $\ssim 0.6$ in the orbital plane and initially
antialigned with each other. We then vary the initial linear
momentum of the holes for fixed binding energy, which is equivalent
to modifying the initial eccentricity. We find that orbital
eccentricity can boost the final recoil by up to $\ssim 25\%$. The
binaries that receive the largest kick of $\ssim 2600$ km/s have
moderate eccentricity $e_t\ssim 0.3$
\cite{Memmesheimer:2004cv,2008PhRvD..78f4069S}. For comparison, the maximal
kick imparted to a quasicircular binary with the same parameters is
$\ssim 2100$ km/s. Our results suggest that the enhanced radiation
of linear momentum is mainly due to the more pronounced asymmetry in the
binary's GW emission rather than the mere consequence of a
larger energy flux.

An additional series of simulations with fixed eccentricity and
varying spin magnitudes allows us to extrapolate these results to
maximally rotating BHs. We predict a maximum superkick of at least
$\ssim 4300$ km/s, compared to the quasicircular result $\ssim 3700$
km/s. We stress that this estimate is conservative because i) we did
not explore the optimal value of the eccentricity as a function of the
spin magnitude and ii) we have constrained the spins to the orbital plane;
partial alignment is known to generate larger recoils
\cite{2011PhRvL.107w1102L,2013PhRvD..87h4027L}. The impact of orbital
eccentricity on these {\em hang-up} kicks with partial spin alignment
is a complex task that we leave for future work: the recoil has a
more complicated dependence on the eccentricity and the initial spin
orientations because of spin precession.

The amplification of superkicks due to orbital eccentricity may
have important consequences for the modeling of GW sources.
For the stellar-mass BHs targeted by ground-based interferometers,
a non-negligible eccentricity at merger would be a powerful signature
of strong and recent interactions with external bodies
(cf.~e.g.~Refs.~\cite{2012ApJ...757...27A,2018PhRvD..97j3014S,2018ApJ...860....5G,2018PhRvD..97j3014S,2018ApJ...855..124S,2019ApJ...871...91Z,2019MNRAS.486.4443F,2019arXiv190711231S}).
If BH binaries coalescing in dynamical environments are indeed
eccentric, our findings further limit the ability of stellar clusters
to retain their merger remnants \cite{2019PhRvD.100d1301G}. For
instance, Refs.~\cite{2018PhRvL.120o1101R,2019PhRvD.100d3027R} found
that dynamical interactions in globular clusters are a viable
formation mechanism to explain multiple generations of eccentric
BH mergers.  The calculation of the retention fraction, however,
does not take into account the significant kick enhancement due to
eccentricity that we have found in this work.  Given the low escape
speed of globular clusters, this amplification may considerably
reduce the predicted number of second-generation BH mergers.

For the case of supermassive BH binaries, eccentric sources are
commonly invoked to explain current PTA limits. Orbital eccentricity
shifts some of the emitted power to higher frequencies, causing a
turnover in the predicted spectrum
\cite{2007PThPh.117..241E,2010ApJ...719..851S,2015PhRvD..92f3010H,2016ApJ...817...70T}.
The presence of this feature allows current astrophysical formation
models calibrated on galaxy counts to more easily accommodate the
measured upper limits. Our work highlights that kicks may be higher
than currently assumed, further reducing the merger rate and the
predicted stochastic GW background.

Numerical-relativity simulations now provide a thorough understanding
of the properties of the BH remnants left behind following mergers of
BHs on quasicircular orbits. Efficient and accurate models for final
mass, spin, and kick are available and routinely implemented in
astrophysical predictions. For eccentric orbits, the additional
dimensionality of the parameter space increases the computational
resources required to accurately predict waveforms and remnant
properties.  Comparatively few numerical studies have focused on the
eccentric regime in the past
\cite{2008PhRvD..77h1502H,2008PhRvD..78f4069S,2010PhRvD..82b4033H},
but more recently, systematic efforts in GW modeling have expanded into
the eccentric regime \cite{Ramos-Buades:2019uvh}. We hope that our
findings have further demonstrated the fertile ground of this class of
binaries and that they will spark future work in this direction.

\section*{Acknowledgments}

We thank V. Baibhav for discussions. 
U.S. is supported by
the European Union’s H2020 ERC Consolidator Grant ``Matter and strong-field
gravity: new frontiers in Einstein's theory'' Grant
No.~MaGRaTh--646597,
and
the STFC Consolidator Grant No. ST/P000673/1.
D.G. is supported by Leverhulme Trust Grant No. RPG-2019-350.
E.B. is supported by
NSF Grant No. PHY-1912550,
NSF Grant No. AST-1841358,
NASA ATP Grant No. 17-ATP17-0225,
and
NASA ATP Grant No. 19-ATP19-0051.
This work has received funding from the European Union's Horizon
2020 research and innovation programme under the Marie Sk\l odowska-Curie
Grant No.~690904.
This work was supported by
the GWverse COST Action CA16104, ``Black holes, gravitational waves and
fundamental physics''.
Computational work was performed on
the SDSC Comet and TACC Stampede2 clusters through
NSF-XSEDE Grant No.~PHY-090003, Cambridge CSD3 system 
through STFC capital Grants No.~ST/P002307/1 and No.~ST/R002452/1, and STFC
operations Grant No.~ST/R00689X/1;
the University of Birmingham \mbox{BlueBEAR} cluster; the Athena cluster at HPC Midlands+ funded by EPSRC Grant No. EP/P020232/1; and the Maryland
Advanced Research Computing Center (MARCC).


\bibliography{newuli}
%


\end{document}